\documentclass[10pt,conference]{IEEEtran}

\usepackage{epstopdf}
\usepackage{graphicx}
\usepackage[
			urlcolor  = blue
		   ]{hyperref}
\usepackage{subcaption}
\usepackage{amsmath,amssymb}
\usepackage{subfloat}
\usepackage{float}

\begin{document}

\title{DARVIZ: Deep Abstract Representation, Visualization, and Verification of Deep Learning Models}

\author{\IEEEauthorblockN{Anush Sankaran, Rahul Aralikatte, Senthil Mani, Shreya Khare, Naveen Panwar, Neelamadhav Gantayat}
	\IEEEauthorblockA{IBM Research, India\\
		\{anussank, rahul.a.r, sentmani, shkhare4, navpanwa, neelamadhav\}@in.ibm.com}
}

\maketitle
\begin{abstract}
Traditional software engineering programming paradigms are mostly object or procedure oriented, driven by deterministic algorithms. With the advent of deep learning and cognitive sciences there is an emerging trend for data-driven programming, creating a shift in the programming paradigm among the software engineering communities. Visualizing and interpreting the execution of a current large scale data-driven software development is challenging. Further, for deep learning development there are many libraries in multiple programming languages such as TensorFlow (Python), CAFFE (C++), Theano (Python), Torch (Lua), and Deeplearning4j (Java), driving a huge need for interoperability across libraries.

We propose a model driven development based solution framework, that facilitates intuitive designing of deep learning models in a platform agnostic fashion. This framework could potentially generate library specific code, perform program translation across languages, and debug the training process of a deep learning model from a fault localization and repair perspective. Further we identify open research problems in this emerging domain, and discuss some new software tooling requirements to serve this new age data-driven programming paradigm.
\end{abstract}


\IEEEpeerreviewmaketitle

\graphicspath{{./}}

\section{Introduction}
Deep learning is a branch of machine learning that deals with extracting features from data in an unsupervised manner. With the help of deep learning, AI and cognition has penetrated a wide variety of applications performing as good as humans or sometimes even better in some challenging tasks such as remembering conversations, perceiving speech, and understanding images~\cite{deepapp}. A deep learning model is essentially a chain of interconnected and predefined layers, with a set of manually defined hyper-parameters for each layer (such as number of nodes, learning rate, size of filters, etc.), and a set of parameters that are learnt using some provided data (such as weights)~\cite{lecun2015deep}. For example consider GoogLenet, a state-of-the-art deep learning model built by Google's DeepMind~\cite{szegedy2015going}. This network consists of 22-layers, where each layer transforms the input data using a transformation weight matrix to perform a classification task. Currently, GoogLeNet model can be implemented and trained in any one of the publicly available platforms such as TensorFlow\footnote{\url{https://www.tensorflow.org/}}, Torch\footnote{\url{http://torch.ch/}}, or CAFFE\footnote{\url{http://caffe.berkeleyvision.org/}}. Despite its increasing success, deep learning algorithms encounter some fundamental software engineering challenges such as model verification, validation, and interpretation. 

\begin{figure}[!t]
	\begin{center}
		\centerline{\includegraphics[width=3.2in]{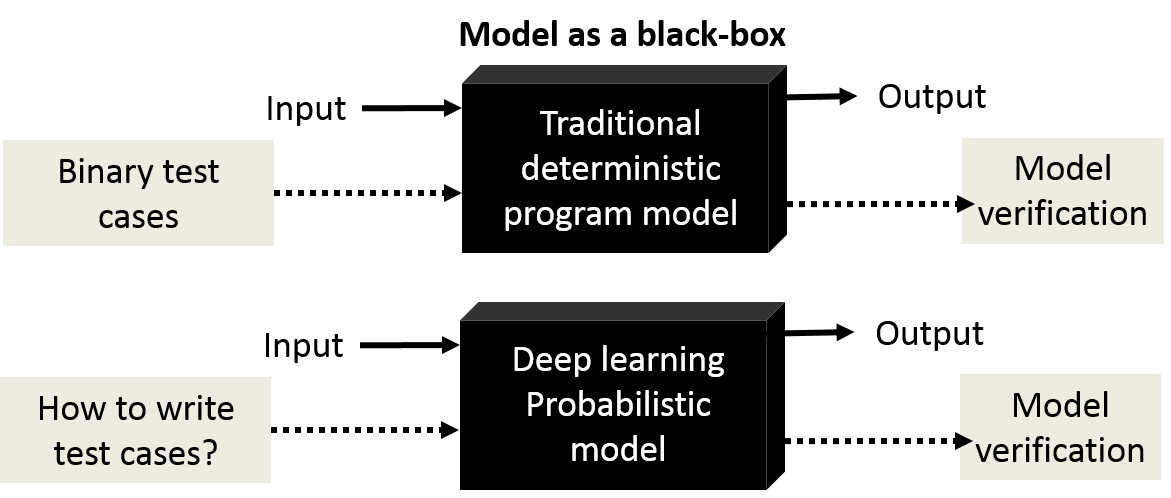}}
		\caption{Comparison of traditional programming and deep learning model designing from the perspective of software engineering model verification and validation.}
		\label{fig:compare}
	\end{center}
	\vspace{-28pt}
\end{figure}

\begin{figure*}[!t]
	\begin{center}
		\centerline{\includegraphics[width=6.7in]{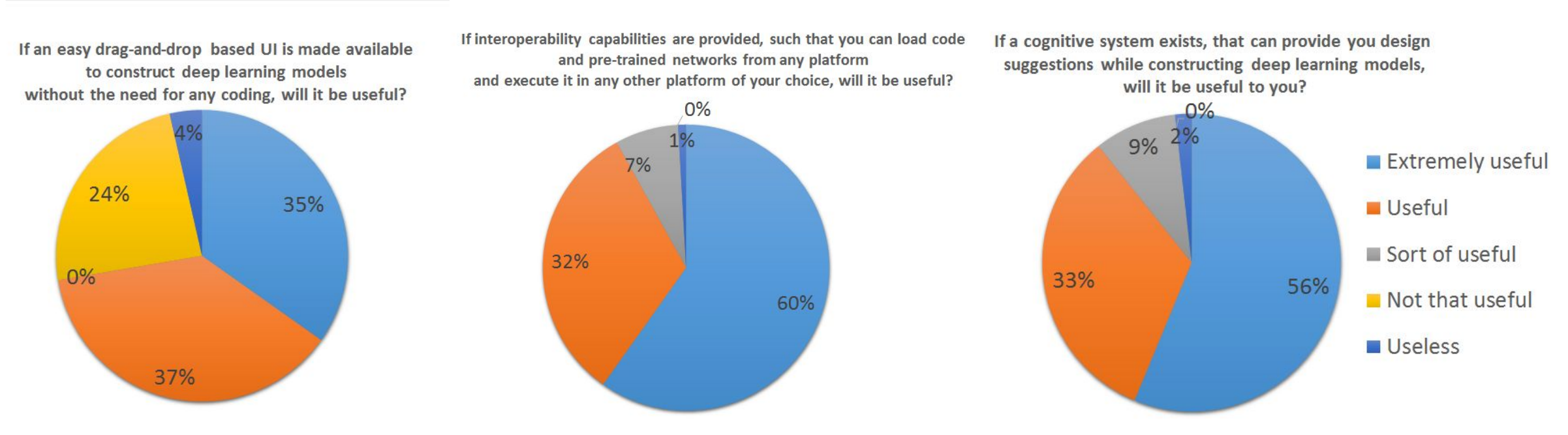}}
		\caption{A summary of user responses obtained from the conducted quantitative survey.}
		\label{fig:cap22}
	\end{center}
	\vspace{-24pt}
\end{figure*}
Traditional software models are deterministic in nature, and hence binary test cases can be written for validating them as shown in Fig.~\ref{fig:compare}. However, as data-driven deep learning models are probabilistic, validating these models becomes highly challenging. Peter Norvig, a pioneer in the field of AI, is an evangelist for this newly emerging research trend of Machine Learning Driven Programming (MLDP)\footnote{In this paper, the term machine learning is used as an umbrella phrase to represent both machine learning and deep learning.} and identifies machine learning as an agile tool for software engineering~\cite{peter}. The ``Software Engineering for Machine Learning" workshop conducted in NIPS 2014, a premium AI conference, emphasizes the growing attention and the need for developing better software engineering tools for proficient development and testing of machine (or deep) learning, rather than the inverse thought process of applying machine learning in software engineering problems. Inspired from the works of Peter Norvig~\cite{norvig2014machine} and Sculley et al.~\cite{43146}, we identify some open ended software engineering challenges in this new paradigm. We conduct a quantitative user survey from more than $100$ software engineers and machine learning researchers, to understand and validate these challenges. We propose a Model-Driven Development (MDD) based solution framework, to serve the data-driven programming paradigm for abstract representation, visualization, and verification of deep learning models.

\section{Quantitative User Survey}
An initial pilot study was conducted with six experts in deep learning  with a purpose of perceiving and understanding the different challenges faced in building deep learning models from a software tooling perspective. Using the pointers from this pilot, a quantitative survey was conducted with $113$ software engineers and researchers from various organizations and varying expertise, demographics, gender, and age. $75\%$ of the responders are good in programming ($4$ or $5$ out of $5$) and $92\%$ of them have at least taken a course in deep learning. The survey form is available live at: \url{https://goo.gl/forms/DA49kKteRyv3Ztgm1} and the results of the survey responses are summarized at: \url{https://goo.gl/dwBYAj}.

From the pilot study, three different scenarios were identified and are provided to the users in this survey to understand and validate the usefulness of the proposed solution framework. The questions corresponding to the three scenarios are: (i) Design: ``How much time does it take for you to implement a research algorithm using a platform that you are comfortable with?", to which almost $83\%$ of the users answered at least a couple of days including failure to successfully implement the algorithm, (ii) Interoperability: ``Other researchers make their code and pre-trained models available in a specific platform they work with (for e.g CAFFE). If you typically work on a different platform (for e.g TensorFlow), have you felt the need for the CAFFE trained model and weights to be made available in TensorFlow as well?", to which $91\%$ of the users agreed that this interoperability is needed, and (iii) Debugging: ``Many times we have trained models only to realize that it has not learnt anything meaningful. Would you agree that it will be useful to have a system which can predict if a model will learn something meaningful or not, fairly early in the training process?", to which more than $87\%$ of the people agreed.

Interestingly, $86\%$ of the users who have rated themselves with the highest rating in `programming ability' have responded that it takes them at least a couple of days to successfully or unsuccessfully implement an existing deep learning model. This indicates that the existing deep learning libraries lack required features for quicker and efficient implementation and prototyping. As a solution, $72\%$ of the responders validated the usefulness of an intuitive drag-and-drop based user interface (Fig.~\ref{fig:cap22}). $92\%$ of the users wanted interoperability tools, that could convert models from one implementation platform to another. Finally, $89\%$ of the users suggested the need of a cognitive system that could suggest hyper-parameters and assist in deep learning model debugging. 

\begin{figure*}[!t]
	\begin{center}
		\centerline{\includegraphics[width=6in]{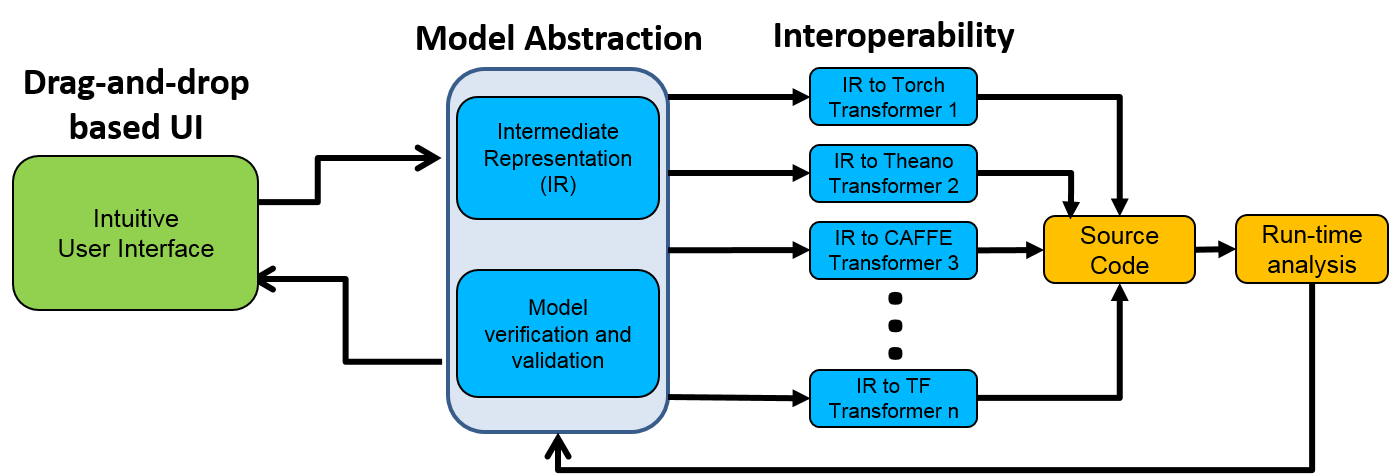}}
		\caption{The proposed solution framework containing the set of new tools that could cater to the needs of the new age of data-driven software development.}
		\label{fig:model}
	\end{center}
	\vspace{-24pt}
\end{figure*}

To capture the additional challenges faced by the responders, an open-ended free form text question was included in the survey. Some highly valuable feedback were obtained including, ``\textit{the lack of proper documentation}", ``\textit{training times are too long}", ``\textit{Implementation of out-of-the-box deep architectures is quite difficult in the available platforms}", ``\textit{Input data processing}", ``\textit{easy visualization (attention) toolkit for model outputs}".
Augmenting our hypothesis with the user survey validations, the following software engineering problems in data-driven programming are identified:
\begin{itemize}
	\item \textbf{Model Validation:} While constructing the layers of a model, there are many design and hyper-parameter choices. Can incorrect design choices be identified and better hyper-parameters be suggested during designing the model?
	\item \textbf{Fault Detection and Debugging:} Due to the large volumes of training data, training deep learning models typically take days or even weeks to converge. Researchers often realize that the training was unsuccessful due to some model fault or incorrect choice of hyper-parameters, only after the entire training is done. Can we perform real-time fault detection and provide debugging capabilities, such that, model faults can be diagnosed much earlier during training?   
	\item \textbf{Model Abstraction and Interoperability:} Current software engineering tools and platforms are ill-suited for design abstraction and model-driven development of deep learning models. Can a model that is implemented in a specific platform (For eg., CAFFE) be abstracted and realized in another platform (For eg., TensorFlow)?
	\item \textbf{Efficient Designer to Improve Productivity:} Implementing deep learning algorithms involves a very steep learning curve and are generally prone to coding and logical errors. Can we build a design tool that provides an intuitive UI for designing deep learning architectures which can also generate code in any of the realization platforms?
\end{itemize}

\section{Proposed Solution Framework}
In this research, we propose an initial solution framework called DARVIZ: Deep Abstract Representation, Visualization, and Verification and is made available at \url{http://darviz.mybluemix.net}. Fig.~\ref{fig:model} shows the three primary aspects of DARVIZ and are detailed as follows:
\begin{enumerate}
	\item ``No-code" intuitive deep model designing,
	\item Model abstraction to provide interoperabilty across platforms, and
	\item Model validation features to be able to interpret what the model has learnt and offer debugging services.
\end{enumerate}
The different features of these three aspects are explained further in detail.

\begin{figure}[!t]
	\begin{center}
		\centerline{\includegraphics[width=3.4in]{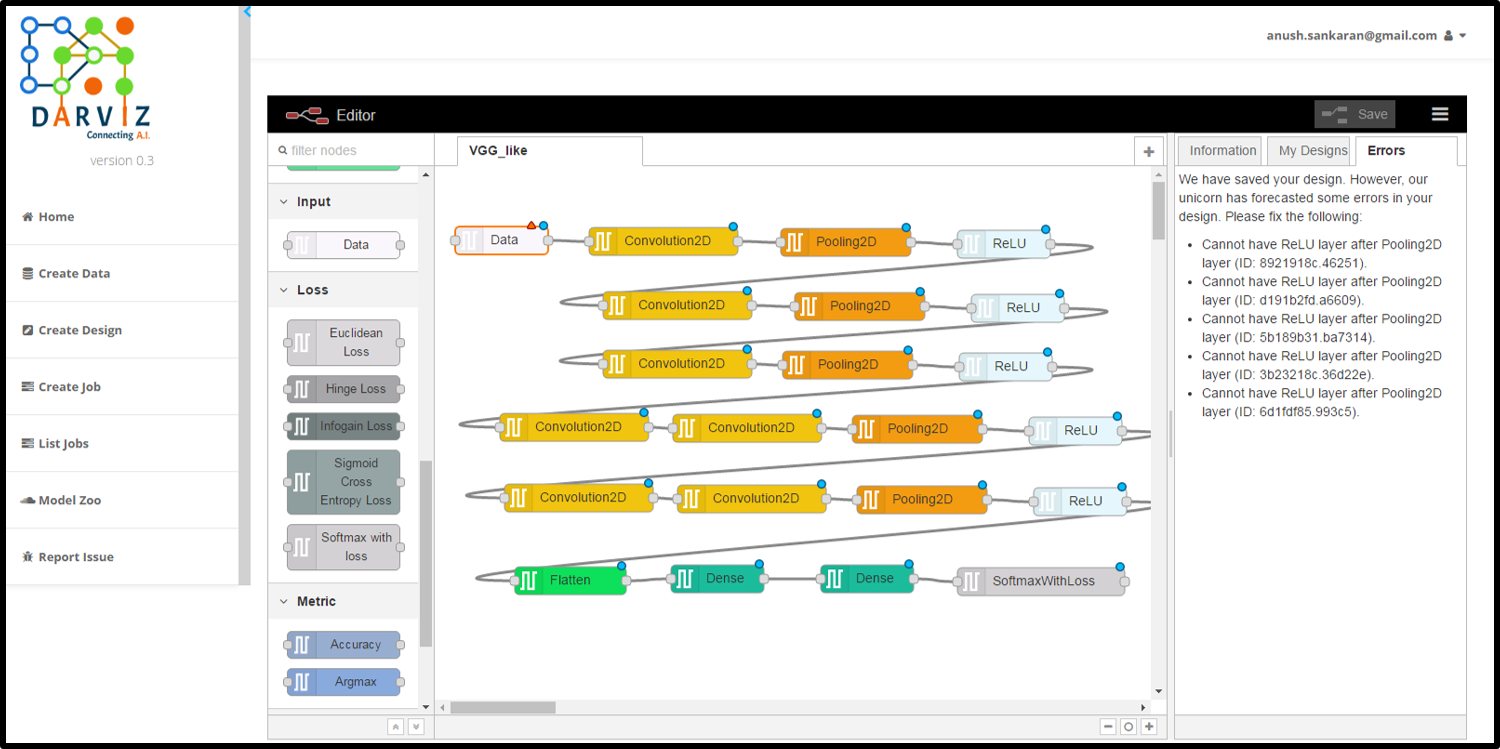}}
		\caption{One of the standard deep learning models in computer vision domain called `VGG network'~\cite{simonyan2014very}, designed using our drag-and-drop UI from scratch.}
		\label{fig:ui}
	\end{center}
	\vspace{-24pt}
\end{figure}

\begin{table*}[!h]
	\begin{center}
		\caption{\label{sumTable2}A comparison of some popular publicly available deep learning platforms and their features.}
		\begin{tabular}{|l|l|l|l|l|}
			\hline
			Platform & Language & Wrapper & Visualization Framework & Challenges \\ \hline 
			
			TensorFlow & Python & Tensorflow-Slim, Keras & TensorBoard & 
			High learning curve of tensor operations \\ \hline
			
			CAFFE & C++ & PyCaffe, MatCaffe
			& NVIDIA DIGITS & 
			Native design scope limited to CNN \\ \hline
			
			Torch & Lua & PyTorch & NVIDIA DIGITS & 
			Proprietary model and designs
			\\ \hline
			
			Theano & Python & Lasagne, Blocks, Keras & - & 
			High learning curve of library definitions
			\\ \hline
			
			DL4J & Java & -
			& - & 
			Poor interoperability with other frameworks
			\\ \hline
			
		\end{tabular}		
	\end{center}
	\vspace{-20pt}
\end{table*}

\subsection{``No-code" Intuitive Designing}
Intuitively, deep learning models are graph like structures having a set of pre-defined layers and hyper-parameters, and driven by data. Implementing these models using a programming language such as Python, C++, or Lua not only takes time and has a learning curve, but also is prone to errors. One feedback received in the user survey is quoted, ``\textit{Many times, for newer research problems, model design choice is not the time consuming part; the formatting and encoding of input data before being fed into an embedding layer of the network takes considerable amount of effort}". This highlights the importance of providing data handling features as part of the UI framework, so that developers are able to seamlessly transition from data munging to deep learning model designing.

The proposed DARVIZ framework consists of independent modules for (i) handling data and performing data preprocessing, (ii) constructing deep learning models from scratch, and (iii) a model zoo, which acts a unified public repository of all the pre-trained models in literature that could be loaded and reused. Data manipulation features in GUI provide the capability for loading data of any format and structure, and performing pre-processing with just a few button clicks. An intuitive drag-and-drop UI will help in construction of new deep learning models as well as quick reproduction of existing models from research papers. For example, consider one of the benchmark deep learning models used in computer vision called the VGG network~\cite{simonyan2014very}. The TensorFlow implementation\footnote{\url{https://github.com/machrisaa/tensorflow-vgg/blob/master/vgg19_trainable.py}} consists of about 400 LOC in Python and typically takes a few hours to few days of development effort. However, the same VGG model was constructed using our drag-and-drop UI, as shown in Fig.~\ref{fig:ui}, in just a few minutes by a developer with minimum domain knowledge. Further, the entire TensorFlow code (or the code in any library) of the designed model can be automatically generated and exported with a single button click.

\subsection{Model Abstraction}
The availability of multiple platforms in multiple programming languages has led to a new challenge of model interoperability across platforms. Consider an example scenario, where researchers have proposed a novel text summarization algorithm using deep learning and have made their code and pre-trained model available in TensorFlow\footnote{\url{https://github.com/tensorflow/models/tree/master/textsum}}. However, if another group of deep learning researchers who are proficient in Torch want to modify this, currently, it is neither possible to import the code nor possible to convert the pre-trained model from TensorFlow to Torch. 

Also, each of these platforms have some advantages and additional features, that the users would like to exploit while building deep learning models. For example, one platform could support more layer definitions while another platform is optimized in handling tensor operations. To address such challenges in DARVIZ, we propose a model abstraction module for both model design and model parameters. Such a module could promote model driven development, where the models already implemented in any of the existing platforms could be converted to the abstract intermediate representation and from the intermediate representation could be realized in any other platform of choice. 

\subsection{Model Validation and Debugging}
One of the most pressing challenges while authoring new deep learning models, is the manual selection of hyper-parameter values and the design choices to be made. Currently, researchers are manually overfitting the model hyper-parameters for given dataset and problem. To make the trained models more generic, there is a lack of debugging tools that could assist in performing root-cause analysis and suggest better choices of hyper-parameters  during training process. Further, in the deep learning community, researchers have put in effort to understand how layers work together and this knowledge can be applied for providing real-time design suggestions. A cognitive fault detection framework could learn the same over a period of time and for a given dataset, could potentially validate the developer's design and help in quick development of a fault free deep learning model.

\section{Literature Study}
There are a few efforts in literature that have, in part, attempted to address some of the challenges identified in this research. As shown in Table~\ref{sumTable2}, the primary focus has been to improve the efficiency of deep learning model designing using GUI frameworks and visualization of some basic parameters during training, while there has been minimum focus towards model validation, debugging, and model abstraction. 

The NVIDIA DIGITS framework\footnote{\url{https://developer.nvidia.com/digits}} has been successful in implementing a ``no-code" interface for loading pre-defined deep learning models. DIGITS also offers visualization of certain watch variables such as training loss and accuracy in terms of graphs. However, they support only a few pre-defined network architectures while designing a model from scratch is not addressed. Further although they support CAFFE and Torch frameworks, there is no conversion service provided to enable interoperability. 

Tensorflow's Tensorboard provides a GUI dashboard for visual summarization of the learnings of the model such as training loss over epochs and the data distribution. They also provide a sophisticated software tool for graphical visualization of pre-defined deep learning models. However, features such as intuitive designing of a model from scratch, supporting models designed using other frameworks, and providing debugging insights based on the visualizations are lacking.

Aetros\footnote{\url{http://aetros.com/}} is a platform to quickly design and train deep learning models from scratch. They provide an intuitive graphical framework to design a specific class of deep learning models called Convolution Neural Networks (CNN) which confines to the ``no-code" development strategy. However, it supports only Keras as an underlying platform and further does not provide any tools for validating and debugging the model.

\section{Conclusion}
Although there have been a few efforts to provide software engineering tools to perform efficient deep learning design, in this work, we identify the lack of a holistic perspective in terms of a model driven development based solution for data-driven programming. We highlight the different software engineering tools envisioned for our DARVIZ framework which could lead to efficient designing of deep learning models for people with varied expertise and also can unify the research efforts happening in the domain of deep learning. An initial version of the DARVIZ is made available here: \url{https://darviz.mybluemix.net} 

\small
\bibliography{deep_mdd}{}
\bibliographystyle{plain}

\end{document}